\documentclass[aps,prl,
,showpacs
,twocolumn
,superscriptaddress
]
{revtex4-2}
\usepackage{newfloat,algcompatible}
\usepackage{etoolbox}
\usepackage{multirow}
\usepackage{graphicx,subfigure,bbm}

\usepackage{amsmath}
\usepackage{tabularx}
\usepackage{amsthm}
\usepackage{bbm}
\usepackage{amssymb}
\usepackage{array,bm}
\usepackage{graphicx}
\usepackage{fancyhdr}
\usepackage{color}
\usepackage{extarrows}
\usepackage{enumerate}
\usepackage{epstopdf}
\usepackage[colorlinks,
            linkcolor=black,
            citecolor=black,
            urlcolor=blue
            ]{hyperref}
\usepackage{natbib}
\usepackage{tikz}
\usetikzlibrary{shapes,arrows}
\usepackage{booktabs}
\usepackage{subfigure}
\usepackage{lineno}
\usepackage{lipsum}
\usepackage{braket}
\usepackage[linesnumbered,ruled]{algorithm2e}

\tikzstyle{decision} = [diamond, draw, fill=blue!20,
    text width=4.5em, text badly centered, node distance=3cm, inner sep=0pt]
\tikzstyle{block} = [rectangle, draw, fill=blue!20,
    text width=10em, text centered, rounded corners, minimum height=4em]
\tikzstyle{line} = [draw, -latex']
\tikzstyle{cloud} = [rectangle, draw,fill=red!20, node distance=7cm,
    minimum height=4em]

\def\(({\left(}
\def\)){\right)}
\def\[[{\left[}
\def\]]{\right]}

\newcommand{\be}{\begin{align}}
\newcommand{\ee}{\end{align}}
\newcommand{\bea}{\begin{eqnarray}}
\newcommand{\eea}{\end{eqnarray}}

\DeclareMathAlphabet{\varmathbb}{U}{bbold}{m}{n}

\begin{document}
\title{Irreversibility Enhances Quantum-Enhanced Markov-Chain Monte Carlo}

\author{Zidong Cui}
\email[Corresponding authors: ]{202411210429@std.uestc.edu.cn}
\altaffiliation{These authors contributed equally}
\affiliation{Institute of Fundamental and Frontier Sciences, University of Electronic Science and Technology of China, Chengdu 611731, China}
\affiliation{School of Physics, University of Electronic Science and Technology of China, Chengdu 611731, China}

\author{Kefan Cao}
\altaffiliation{These authors contributed equally}
\affiliation{Institute of Fundamental and Frontier Sciences, University of Electronic Science and Technology of China, Chengdu 611731, China}
\affiliation{Yingcai Honor College, University of Electronic Science and Technology of China, Chengdu 611731, China}

\author{Jiale Chen}
\affiliation{Institute of Fundamental and Frontier Sciences, University of Electronic Science and Technology of China, Chengdu 611731, China}

\author{Lei Wang}
\affiliation{Beijing National Laboratory for Condensed Matter Physics and Institute of Physics, Chinese Academy of Sciences, Beijing 100190, China}

\author{Ying Tang}
\email[Corresponding authors: ]{jamestang23@gmail.com}
\affiliation{Institute of Fundamental and Frontier Sciences, University of Electronic Science and Technology of China, Chengdu 611731, China}
\affiliation{School of Physics, University of Electronic Science and Technology of China, Chengdu 611731, China}
\affiliation{Key Laboratory of Quantum Physics and Photonic Quantum Information, Ministry of Education, University of Electronic Science and Technology of China, Chengdu 611731, China}
\affiliation{Non-classical Information Science Basic Discipline Research Center of Sichuan Province, University of Electronic Science and Technology of China, Chengdu 611731, China}

\begin{abstract}
Detailed balance underlies conventional Markov-chain Monte Carlo (MCMC) algorithms. Yet in classical systems, breaking detailed balance generates irreversible probability currents and can accelerate sampling. Whether irreversibility can similarly enhance quantum MCMC remains an intriguing question. Here we show that irreversibility provides a new route to improving the quantum-enhanced MCMC (QEMC), which combines quantum proposals with classical acceptance. By integrating quantum-generated candidate paths with allocations of probability flows, we develop an irreversible quantum-enhanced Monte Carlo (IQEMC). On spin-glass benchmarks, IQEMC improves the spectral gap over QEMC with a polynomial scaling advantage as the system size increases. In addition to numerically demonstrating faster convergence of magnetization estimates, we verify this improvement experimentally on a quantum processor. The results establish irreversibility as a physically grounded mechanism for enhancing quantum MCMC.
\end{abstract}

\maketitle
\textit{\label{sec:level1}Introduction} -- Detailed balance constitutes the foundation of conventional Markov-chain Monte Carlo, such as the widely used Metropolis-Hastings algorithm~\cite{metropolis1953equation,hastings1970monte}. While detailed balance guarantees convergence to the target distribution, it is not necessary for preserving the stationary distribution, and breaking it generates irreversible probability currents that accelerate sampling. An early study introduced non-detailed-balance MCMC through local detailed balance~\cite{10.1063/1.477973}, which is sufficient to ensure stationarity. Subsequent works developed diverse irreversible MCMC schemes~\cite{ma2015complete,ao2013dynamical,TURITSYN2011410} and demonstrated accelerated convergence in a wide range of systems, including hard-sphere systems~\cite{PhysRevE.80.056704}, Potts models~\cite{suwa}, self-avoiding walks~\cite{Hao_Hu_120503}, dense liquids~\cite{ghimenti2024irreversible1}, hard-disk glasses~\cite{ghimenti2024irreversible2}, and Ising machines~\cite{xu2025combinatorial}. A rigorous analysis showed that violating detailed balance improves sampling efficiency by modifying the spectrum of the transition matrix~\cite{vdbc}. These findings in classical systems raise the question of whether irreversibility can similarly enhance Monte Carlo sampling in the quantum regime.

Addressing this question requires a quantum framework in which irreversibility can be meaningfully incorporated. Alongside quantum MCMC based on quantum random walks~\cite{szegedy2004quantum,richter2007quantum,montanaro2015quantum,lemieux2020efficient,arunachalam2022simpler,preskill2018quantum}, adiabatic state generation~\cite{wocjan2008speedup}, and quantum simulated annealing~\cite{yung2012quantum}, the recent quantum-enhanced Monte Carlo (QEMC)~\cite{nature2023qe} achieves a polynomial speedup in sampling the Boltzmann distribution relative to classical methods. More recently, fully coherent quantum MCMC schemes based on Hamiltonian-generated quantum proposals have been developed to improve quantum sampling~\cite{incudini2026practical}. Since QEMC is compatible with noisy intermediate-scale quantum devices, it provides a practical platform for investigating the role of irreversibility. Despite many subsequent efforts to improve QEMC such as by annealing~\cite{arai2025quantum} and neural-network surrogates~\cite{9nhx-5pym}, these approaches entail greater computational complexity, while the role of irreversibility remains unexplored. It is thus unclear whether the advantages of irreversible sampling observed in classical MCMC can be realized in QEMC.

\begin{figure}[!htbp]
{\includegraphics[width=1\linewidth]{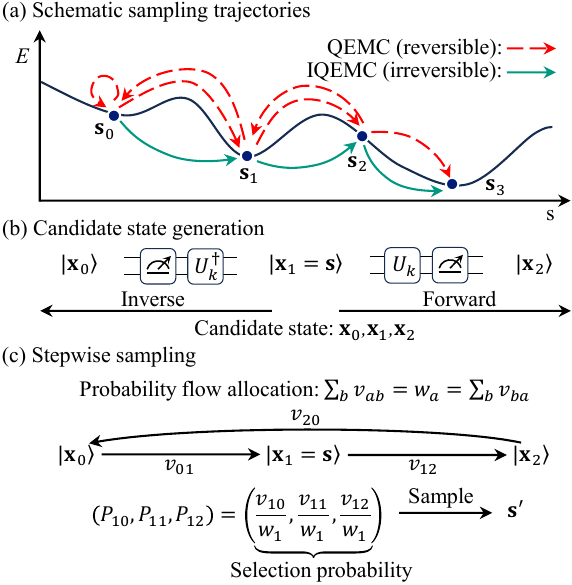}}
\caption{Schematic of IQEMC. (a) Convergence behavior of QEMC and IQEMC. IQEMC reduces trapping and revisits, enabling faster exploration of distinct states $\mathbf{s}_a$ and accelerating convergence to the target distribution. (b) IQEMC generates candidate states $\mathbf{x}$ for the current state $\mathbf{s}$ using randomly selected forward ($U_k$) and inverse ($U_k^\dagger$) circuits. (c) IQEMC samples the next state $\mathbf{s}'$ according to the transition probability $P$, determined by the allocated probability flow $v_{ab}\equiv v_A({\mathbf{s}}_a,{\mathbf{s}}_b)$ and the candidate's weight $w_a\equiv w_A({\mathbf{s}}_a)$.
}
\label{figure1}
\end{figure}

A central challenge for this realization is to allocate probability flows that preserve the target distribution in the quantum regime, where the global balance condition requires equal total incoming and outgoing flows at each state. For quantum proposals, this task is complicated by the difficulty of efficiently evaluating proposal probabilities. Inspired by classical probability-flow allocation~\cite{suwa,ghimenti2024irreversible2}, we address this challenge by developing an irreversible quantum-enhanced Monte Carlo (IQEMC) method. IQEMC employs a forward-inverse quantum path design (Fig.~\ref{figure1}) to enable irreversible probability flows while preserving global balance. This design avoids the explicit evaluation of quantum proposal probabilities and maintains quantum circuit complexity similar to QEMC on quantum hardware, requiring only that the number of executions of the fixed circuits scales linearly with the path length.

The advantage of IQEMC is reflected in the improved convergence efficiency. Benchmarks on spin-glass models show that IQEMC achieves a polynomial advantage in spectral-gap scaling over QEMC. This advantage persists even when QEMC is given matched computational resources (End Matter). We also experimentally verify on a quantum processor that IQEMC achieves faster convergence of magnetization estimates than QEMC, consistent with the numerical simulations. We further introduce a time-dependent variant, $t$-IQEMC, which preserves the polynomial scaling advantage and yields larger spectral gaps. These results identify irreversibility as a new route to enhancing quantum MCMC.

\textit{\label{sec:level2.1}Quantum-enhanced MCMC} -- The QEMC~\cite{nature2023qe} algorithm enables efficient sampling from Boltzmann distributions, exhibiting enlarged spectral gaps and polynomial speedups over classical MCMC methods at low temperatures $T$. We consider sampling from an Ising model with state $\mathbf{s}=\{s_l\}_{l=1}^{n}$, where $s_l\in\{-1,+1\}$, and energy

\begin{align} 
E(\mathbf{s})=-\sum_{ l<m}J_{lm}s_ls_m-\sum_l h_l s_l, 
\end{align}
and the Boltzmann distribution $\pi(\mathbf{s})\propto e^{- E(\mathbf{s})/T}$. Here $J_{lm}$ are pairwise couplings, $h_l$ are external fields, and $T$ is the temperature~\cite{lucas2014ising}. Encoding $\mathbf{s}$ into a quantum state $\ket{\mathbf{s}}$, sampling is implemented via quantum evolution $U=e^{-iHt}$ with $H=(1-\gamma_{0})\alpha H_{\text{prob}}+\gamma_{0} H_{\text{mix}}$, where $\gamma_{0}$ is uniformly sampled from the interval $[0.6,0.8]$~\cite{nature2023qe} and $\alpha=\sqrt{n/(\sum_{1\leq l<m\leq n}J_{lm}^{2}+\sum_{l=1}^{n}h_{l}^{2})}$. For the state space $\mathcal{C}$, the ``problem'' Hamiltonian $H_{\text{prob}}=\sum_{\mathbf{s}}E(\mathbf{s})\ket{\mathbf{s}}\bra{\mathbf{s}}, \mathbf{s}\in \mathcal{C}$ encodes the target distribution, while $H_{\text{mix}}=\sum_l X_l$ induces transitions between states. This dynamics induce long-range transitions between near-degenerate states ($\Delta E\approx0$), yielding symmetric proposals $|\braket{\mathbf{s}'|U|\mathbf{s}}|^2=|\braket{\mathbf{s}|U|\mathbf{s}'}|^2$ with detailed balance $P(\mathbf{s}|\mathbf{s}')\pi(\mathbf{s}')=P(\mathbf{s}'|\mathbf{s})\pi(\mathbf{s})$, under the Metropolis rule $f_{MH}(\mathbf{s}'|\mathbf{s})=\min\{1,e^{-\Delta E/T}\}$ with $\Delta E=E(\mathbf{s}')-E(\mathbf{s})$. The convergence rate is given by the absolute spectral gap~\cite{jerison2013general}:
\begin{align}
   \delta = 1-\max(|\lambda|),
\end{align}
where $\lambda$ denotes an eigenvalue of the transition matrix associated with a nonstationary mode; small $\delta$ leads to slow mixing in metastable landscapes~\cite{mcmt}. This spectral measure is applicable to both reversible and irreversible finite-state chains. Despite these favorable properties, QEMC remains suboptimal when initialized at high energies~\cite{orfi2024bounding,orfi2024barriers}. Motivated by classical results on accelerated sampling via broken detailed balance~\cite{vdbc,suwa}, we provide an algorithm without detailed balance.

\textit{\label{sec:level2.2} Irreversible quantum-enhanced Monte Carlo} -- IQEMC relaxes the detailed-balance condition through the allocation of probability flows (Fig.~\ref{figure1}), enabling faster exploration of distinct states and accelerating convergence to the target distribution. IQEMC generates candidate states using a fixed bank of $K$ unitary circuits $\{U_k\}_{k=1}^{K}$ (End Matter). For each $k$, both $U_k$ and $U_k^\dagger$ are state-independent. At one Monte Carlo step, we independently and uniformly sample a circuit index $k\in\{1,\ldots,K\}$ and an anchor position $\xi$, and set $\mathbf{x}_{\xi}=\mathbf{s}, \xi\in\{0,\dots ,n\}$. An ordered path $\mathbb{Z}=(\mathbf{x}_{0},\dots,\mathbf{x}_{z},\dots,  \mathbf{x}_{n})$ is generated by applying $U_k$ along increasing indices and $U_k^\dagger$ along decreasing indices. The details of inverse circuits, circuit selection, and random anchors are in the Supplemental Material. Each circuit application is followed by computational-basis measurement, and the measured state initializes the next application. The forward and inverse circuits yield the same transition probability $q_{k}(\mathbf{x}_{z},\mathbf{x}_{z+1})=|\langle \mathbf{x}_{z+1}|U_k|\mathbf{x}_{z}\rangle|^2=|\langle \mathbf{x}_{z}|U_k^\dagger|\mathbf{x}_{z+1}\rangle|^2$. The probability of generating the auxiliary object $A=(k,\mathbb{Z})$ from $\mathbf{s}$ is
\begin{equation}
p(A\mid\mathbf{s})=\frac{c_A(\mathbf{s})}{K(n+1)}B_A,
B_A=\prod_{z=0}^{n-1}q_k(\mathbf{x}_{z},\mathbf{x}_{z+1}),
\label{eq:path-generation}
\end{equation}
where $c_A(\mathbf{s})=\sum_{z=0}^{n}\mathbf{1}_{\{\mathbf{x}_{z}=\mathbf{s}\}}$ counts the occurrences of $\mathbf{s}$ along the path $\mathbb{Z}$.

 Let $\mathcal C_A$ denote the set of distinct states on the path. The path-conditioned  probability is
\begin{equation}
 \rho_A(\mathbf{s})=\frac{\pi(\mathbf{s})p(A\mid \mathbf{s})}{\mu(A)}
 =\frac{w_A(\mathbf{s})}{W_A},
 \label{eq:conditional-weights}
\end{equation}
where $ w_A(\mathbf{s})=c_A(\mathbf{s})e^{-E(\mathbf{s})/T}$ is the candidate weight, $\mu(A)=\sum_{\mathbf{s}\in \mathcal{C}}\pi(\mathbf{s})p(A\mid \mathbf{s})$ is the path probability, and $W_A=\sum_{\mathbf{s}\in\mathcal C_A}w_A(\mathbf{s})$ is the total candidate weight. The common quantum factor $B_A$ cancels, so the allocation requires energies and multiplicities, but no explicit evaluation of quantum proposal probabilities.

In IQEMC, both the allocation of probability flows and the transition probability matrix are constructed from these weights. Sort the distinct candidates by placing the state with the largest weight first, followed by the others in increasing order of their indices (defined in the Supplemental Material). Denote the resulting weights by $\{w_A({\mathbf{s}}_{a})\}_{a=0}^{|\mathcal C_A|-1}$. If multiple candidates share the maximum weight, the one with the smallest state index is placed first. The allocation of probability flows is
    \begin{equation}
    \begin{aligned}
v_A({\mathbf{s}}_{a}&,{\mathbf{s}}_{b})=\sum_{g=0}^{1}\Big[\min\!\bigl(
        S_{A,a}+w_{A}({\mathbf{s}}_{0}),\,S_{A,b}+gW_A
        \bigr)\\
        &-\max\!\bigl(
        S_{A,a-1}+w_{A}({\mathbf{s}}_{0}),\,S_{A,b-1}+gW_A
        \bigr)\Big]_{+},
    \end{aligned}
\end{equation}
where $S_{A,a}=\sum_{j=0}^{a}w_{A}({\mathbf{s}}_{j})$, $S_{A,-1}=0$, and $[\cdot]_{+}=\max(\cdot,0)$. This allocation satisfies $ \sum_b v_A({\mathbf{s}}_{a},{\mathbf{s}}_{b})=w_A({\mathbf{s}}_{a})=\sum_b v_A({\mathbf{s}}_{b},{\mathbf{s}}_{a})$. The next state is sampled according to 
\begin{equation}
 P_A({\mathbf{s}}_{a},{\mathbf{s}}_{b})=\frac{v_A({\mathbf{s}}_{a},{\mathbf{s}}_{b})}{w_A({\mathbf{s}}_{a})}.
 \label{eq:conditional-update}
\end{equation}
The allocation of probability flows directly determines the next-state distribution, eliminating the need for an additional Metropolis-Hastings acceptance step. The transition kernel is $P(\mathbf{s}'|\mathbf{s})=\sum_{A:\,p(A\mid\mathbf{s})>0} p(A\mid \mathbf{s})P_A(\mathbf{s},\mathbf{s}')$. We consider only auxiliary paths with $\mu(A)>0$. For $\mathbf{s}'\notin\mathcal C_A$, the conditional probability $\rho_A(\mathbf{s}')$ and all transitions from candidates in $\mathcal C_A$ to $\mathbf{s}'$ are zero. The conditional update preserves $\rho_A$, since
\begin{equation}
\sum_{\mathbf{s}\in\mathcal C_A}\rho_A(\mathbf{s})P_A(\mathbf{s},\mathbf{s}')
=
\frac{1}{W_A}\sum_{\mathbf{s}\in\mathcal C_A}v_A(\mathbf{s},\mathbf{s}')
=
\rho_A(\mathbf{s}').
\label{eq:preserve}
\end{equation}
Consequently, Eqs.~\eqref{eq:conditional-weights}--\eqref{eq:preserve} imply
\begin{equation}
\begin{aligned}
\sum_{\mathbf{s}\in\mathcal{C}}\pi(\mathbf{s})P(\mathbf{s}'\mid\mathbf{s})
&=\sum_A\mu(A)
  \sum_{\mathbf{s}\in\mathcal C_A}\rho_A(\mathbf{s})P_A(\mathbf{s},\mathbf{s}')\\
&=\sum_A\mu(A)\rho_A(\mathbf{s}')\\
&=\pi(\mathbf{s}')\sum_A p(A\mid \mathbf{s}')
=\pi(\mathbf{s}').
\end{aligned}
\end{equation}
Thus, the update preserves the Boltzmann distribution, as numerically verified in the End Matter. Details of the proof are provided in the Supplemental Material. The chain converges to the Boltzmann distribution if $P$ is irreducible and aperiodic. Since the allocated flows can be asymmetric, the resulting dynamics can violate detailed balance. Specifically, the irreversibility of the marginal chain requires
\begin{equation}
\pi(\mathbf{s})
P(\mathbf{s}'\mid\mathbf{s})
\neq
\pi(\mathbf{s}')
P(\mathbf{s}\mid\mathbf{s}')
\end{equation}
for at least one pair $(\mathbf{s},\mathbf{s}')$, and it is not guaranteed for every parameter choice. 

\begin{figure}[htbp]
    \centering
    \includegraphics[width=1\linewidth]{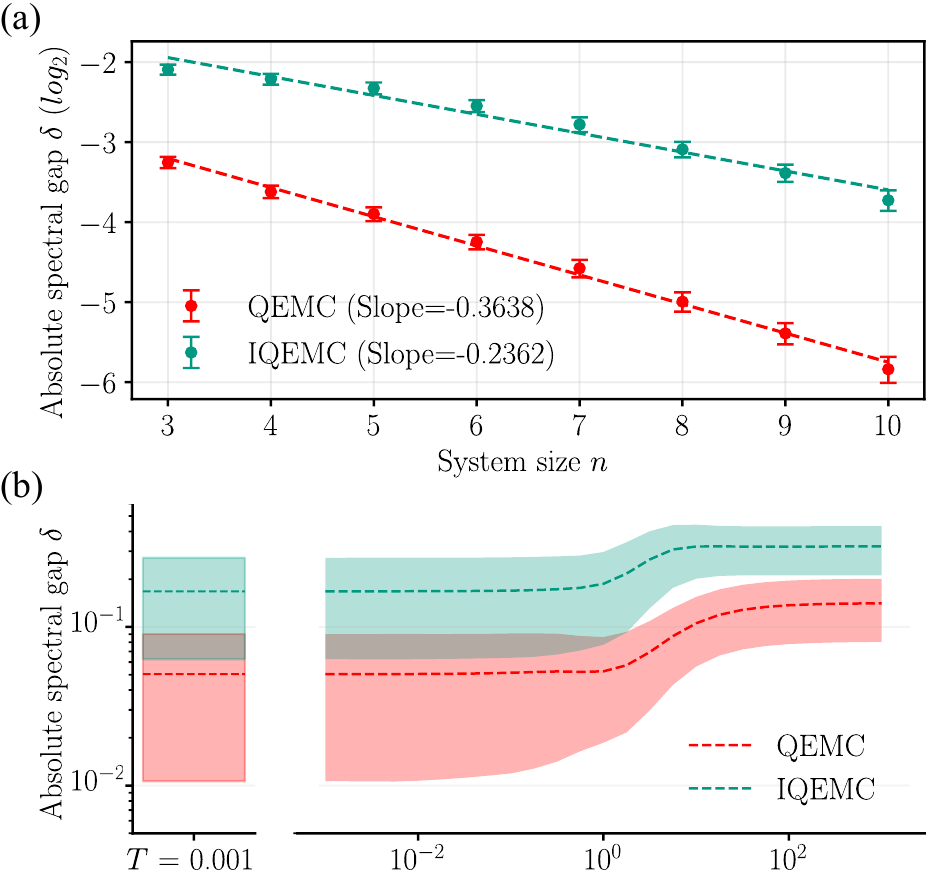}
    \caption{Average absolute spectral gap of IQEMC and QEMC in $500$ fully connected random Ising models. (a) The spectral gap of IQEMC increases relative to QEMC with system size  ($T=0.1$), indicating a polynomial scaling advantage. (b) Spectral gap versus temperature $T$ for $n=6$. IQEMC exhibits a larger spectral gap across the temperature range.}
    \label{fig:Main_result}
\end{figure}

\textit{\label{sec:level3.1} Average convergence efficiency} -- We benchmark the spectral gap $\delta$ to demonstrate the efficiency of IQEMC in the fully connected random Ising model: 

\begin{align}
    H_{\text{prob}}=-\sum_{l< m} J_{lm} Z_l Z_m - \sum_l h_l Z_l,
\end{align}
where $J_{lm}$ and $h_l$ are drawn from standard normal distributions. For system sizes $3\leq n\leq 10$, IQEMC achieves a larger $\delta$ than QEMC~\cite{nature2023qe}, with the advantage increasing with system size (Fig.~\ref{fig:Main_result}a). Results are averaged over $500$ instances. Moreover, IQEMC exhibits a polynomial scaling advantage in $\delta$ over QEMC, as indicated by the fitted scaling of $\log_2\delta$ with system size $n$. Fig.~\ref{fig:Main_result}b shows a consistent advantage across temperatures for $n=6$. Both IQEMC and $t$-IQEMC exhibit a polynomial advantage. A comparison between IQEMC and classical MCMC is provided in the Supplemental Material~\cite{supplementary}.

\begin{figure}[!htbp]
{\includegraphics[width=1\linewidth]{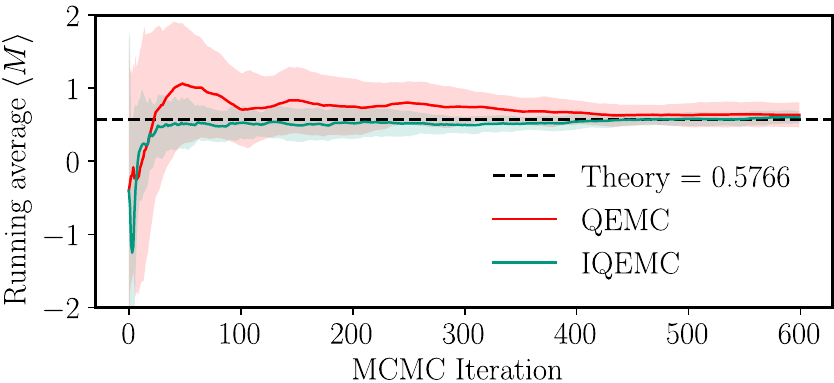}}
\caption{Convergence of the mean total magnetization in numerical simulations. Results are averaged over $10$ independent sampling trajectories for a single fully connected random Ising instance ($n=6$, $T=0.1$). IQEMC (green) converges to the exact reference value in fewer steps than QEMC (red).}
\label{fig3}
\end{figure}

\textit{\label{sec:level3.2}Magnetization estimation} -- While the spectral gap characterizes the intrinsic convergence properties of the Markov chain, practical applications require accurate estimation of physical observables. We therefore estimate the mean total magnetization $\braket{M}$ in the Ising model with respect to the Boltzmann distribution $\pi(\mathbf{s})$ as $\braket{M} = \sum_{\mathbf{s}} \pi(\mathbf{s}) M(\mathbf{s})$, where the magnetization of state $\mathbf{s}$ is $M(\mathbf{s}) = \sum_{l=1}^n s_l$. In Fig.~\ref{fig3}, IQEMC initially deviates more from the theoretical value than QEMC because it explores a broader state space. Overall, IQEMC exhibits faster convergence than QEMC, consistent with the spectral-gap analysis.

\begin{figure}[htbp]
{\includegraphics[width=1\linewidth]{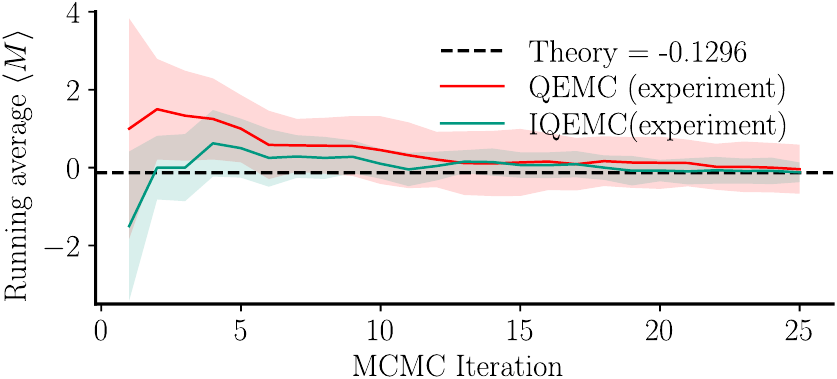}}
\caption{Convergence of the total magnetization tested on the quantum processor. For an $n=3$ fully connected random Ising model at $T=10$, IQEMC (green) converges to the exact reference value in fewer steps than QEMC (red).}
\label{fig_exp}
\end{figure}

\textit{\label{sec:level3.3}Magnetization estimation on quantum devices} -- To demonstrate our algorithm on real quantum devices, we compare QEMC and IQEMC on a fully connected $n=3$ Ising instance at $T=10$ using the AGate-100 quantum processor through the LogicalQubit cloud platform. Both methods use the same circuit bank and initial state. QEMC applies a Metropolis-Hastings update to a single quantum proposal. In contrast, IQEMC generates a four-position path using forward and inverse circuits, then selects the next state through the allocation of probability flows. We monitor the mean total magnetization $\braket{M}$ in the Ising model. The magnetization is measured after each quantum processor execution, so the initial value corresponds to the first sampling step, explaining the initial mismatch between the two curves. The reference value is obtained by exact enumeration of the Boltzmann distribution. In Fig.~\ref{fig_exp}, curves show the mean running average over independent chains, with shaded bands indicating one standard deviation across chains. For this instance, IQEMC approaches the equilibrium magnetization in fewer steps than QEMC, demonstrating faster relaxation of this observable.

\textit{\label{sec:level4} Conclusion} --  We develop an irreversible quantum-enhanced MCMC framework that accelerates convergence by breaking detailed balance. IQEMC achieves a polynomial improvement in spectral-gap scaling over QEMC, even against QEMC under matched quantum resources. Quantum hardware experiments demonstrate faster convergence of IQEMC for physical observables such as magnetization, consistent with numerical simulations. The circuit complexity of IQEMC on quantum processors is similar to that of QEMC, requiring only that the number of executions of the fixed circuits scales linearly with the path length. Time-dependent control further enlarges the spectral gap in $t$-IQEMC. IQEMC identifies nonequilibrium driving as a practical resource for quantum sampling while requiring the same circuit depth as QEMC on a quantum processor. Several directions merit further investigation. One is the optimal design of irreversibility~\cite{lu2026steering} to maximize quantum sampling efficiency. Jointly optimizing circuit selection, path length, and probability-flow allocation can improve mixing under a fixed quantum-resource budget. Another direction is extending quantum-hardware experiments to larger systems to assess the scalability of the improvements.

\textit{Acknowledgments} -- We thank Luming Duan for discussions. This work is supported by Projects 12322501, 12575035 of the National Natural Science Foundation of China, and 2026NSFSCZY0124 of the Natural Science Foundation of Sichuan Province. The computational work is supported by the Center for HPC, University of Electronic Science and Technology of China. 

\textit{Code availability} -- The code implementation will be available upon acceptance of the manuscript.

\bibliography{bib}

\appendix
\section*{End Matter}
\twocolumngrid

\begin{table*}[t]
\caption{
Verification of Boltzmann stationarity and irreversibility in IQEMC and $t$-IQEMC. All quantities are computed from the full transition matrix $P$ for a fully connected random Ising instance with $n=5$ at $T=10$. Here, $\pi$ and $\pi_P$ denote the target Boltzmann and independently computed stationary distributions, respectively, and $\mathcal{J}_{\max}$ is the maximum absolute stationary probability current. Both methods preserve the target distribution to numerical precision, violate detailed balance, and exhibit larger absolute spectral gaps than QEMC.
}
\label{tab:exact-balance-n5}
\begin{ruledtabular}
\begin{tabular}{ccccc}
Method
& $\|\pi^{\mathsf T}P-\pi^{\mathsf T}\|_1$
& $\frac12\|\pi_P-\pi\|_1$
& $\mathcal{J}_{\max}$& $\delta$\\
\hline
QEMC
& $1.32\times10^{-16}$
& $7.15\times10^{-16}$
& $2.60\times10^{-18}$
& $0.10158$ \\
IQEMC& $9.82\times10^{-15}$
& $9.63\times10^{-15}$
& $2.11\times10^{-3}$
& $0.27803$ \\
$t$-IQEMC& $3.55\times10^{-15}$
& $1.39\times10^{-15}$
& $2.77\times10^{-3}$
& $0.69251$ \\
\end{tabular}
\end{ruledtabular}
\end{table*}

\textit{Construction of the quantum circuit bank} --  For each Ising instance, we construct a fixed bank of $K$ unitary operators $\{U_k\}_{k=1}^{K}$. The operators depend on the problem instance but not on the current state. We use units with $\hbar=1$ and define
\begin{equation}
H_{\mathrm{prob}}
=
\sum_{\mathbf{s}\in\mathcal C}
E(\mathbf{s})|\mathbf{s}\rangle\langle\mathbf{s}|,
\qquad
H_{\mathrm{mix}}=\sum_{l=1}^{n}X_l,
\end{equation}
For each circuit index $k$, we independently draw $\gamma_{0,k}\sim\mathrm{Unif}[0.6,0.8]$ and $t_k\sim\mathrm{Unif}[2,20]$. The parameter pairs are drawn independently across circuit indices and remain fixed throughout the simulation. The two circuit families are paired using the same random draws: (i) static circuits; (ii) time-dependent circuits. 

For the paired static QEMC circuit, we set $\gamma_k=0.25+{0.35}/{0.2}(\gamma_{0,k}-0.6)$ and use the same $t_k$. The static circuit used for IQEMC is
\begin{equation}
U_k
=
\exp\!\left[-it_k H_k\right],
H_k=(1-\gamma_k)H_{\mathrm{prob}} +\gamma_k H_{\mathrm{mix}}.
\label{eq:static-bank}
\end{equation}

For the alternative circuit bank used in $t$-IQEMC, the sweep rates are assigned deterministically as
\begin{equation}
\kappa_k=
\begin{cases}
-0.001, & k\equiv1\pmod{3},\\
-0.05,  & k\equiv2\pmod{3},\\
-0.1,   & k\equiv0\pmod{3}.
\end{cases}
\end{equation}
For a given circuit, $\kappa_k$ remains constant throughout the evolution. The mixing parameter $\gamma_k(\tau)=\gamma_{0,k}+\kappa_k\tau$ varies with the schedule coordinate $\tau$. The corresponding Hamiltonian is
\begin{equation}
H_k(\tau)
=
[1-\gamma_k(\tau)]H_{\mathrm{prob}}
+
\gamma_k(\tau)H_{\mathrm{mix}},
\end{equation}
The scan is truncated at $\tau_k^{\mathrm{stop}}=\min\left\{t_k,\frac{\gamma_{0,k}-0.1}{-\kappa_k}\right\}$, corresponding to either reaching $t_k$ or reducing the mixing parameter to $0.1$. The time-dependent evolution is implemented using a piecewise-constant midpoint approximation with $\Delta\tau=0.5$. The resulting unitary is a chronologically ordered product of the segment propagators. Thus, one path update uses $n$ complete circuit applications, all sharing the same selected circuit index.

\begin{figure}[!htbp]
    \centering
    \includegraphics[width=1\linewidth]{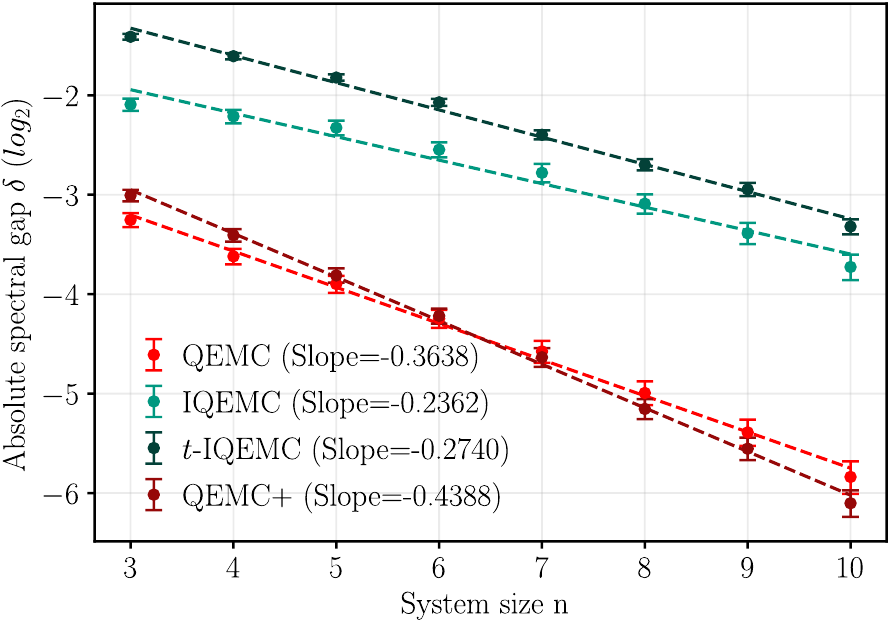}
    \caption{Spectral-gap scaling with system size for the fully connected random Ising model ($T=0.1$). IQEMC and $t$-IQEMC achieve the largest gaps, with IQEMC (green) showing the most favorable scaling and $t$-IQEMC (dark green) a larger absolute gap. QEMC given matched quantum resources (QEMC+, dark red) shows no significant difference from QEMC (red).}
    \label{fig:End_Matter_result}
\end{figure}

\textit{Verification of global balance and irreversibility} -- We verify the stationary distribution and irreversibility of the implemented updates by explicitly constructing the full transition matrices for IQEMC and $t$-IQEMC in a fully connected random Ising model with $n=5$ at $T=10$, using $K=10$ circuits for both methods. First, we evaluate the global-balance residual
\begin{equation}
\epsilon_{\mathrm{bal}}=\left\|\boldsymbol{\pi}^{\mathsf T}P-\boldsymbol{\pi}^{\mathsf T}\right\|_1,
\end{equation}
where $\boldsymbol{\pi}$ is the normalized Boltzmann distribution. This quantity measures whether a single update preserves the target distribution. Independently, we obtain the stationary distribution $\boldsymbol{\pi}_P$ from $P^{\mathsf T}\boldsymbol{\pi}_P=\boldsymbol{\pi}_P$ under normalization, without imposing the Boltzmann distribution.

The discrepancy between the computed stationary distribution and the target distribution is quantified by
\begin{equation}
d_{\mathrm{TV}}
=
\frac{1}{2}
\left\|
\boldsymbol{\pi}_P-\boldsymbol{\pi}
\right\|_1.
\end{equation}
As shown in Table~\ref{tab:exact-balance-n5}, both diagnostics are below $10^{-14}$ for IQEMC and $t$-IQEMC, consistent with floating-point precision. These results numerically confirm that the implemented updates preserve the intended Boltzmann distribution.

To distinguish global balance from detailed balance, we calculate the maximum stationary probability current
\begin{equation}
\mathcal{J}_{\max}
=
\max_{\mathbf{s},\mathbf{s}'}
\left|
\pi(\mathbf{s})P(\mathbf{s}'\mid\mathbf{s})
-
\pi(\mathbf{s}')P(\mathbf{s}\mid\mathbf{s}')
\right|.
\end{equation}
For QEMC, $\mathcal{J}_{\max}=2.60\times10^{-18}$, consistent with detailed balance. In contrast, IQEMC and $t$-IQEMC yield $\mathcal{J}_{\max}=2.11\times10^{-3}$ and $2.77\times10^{-3}$, respectively, well above numerical precision and confirming irreversibility for the tested instance. Together with the negligible global-balance residuals, these results show that nonzero pairwise probability currents can coexist with balanced total incoming and outgoing flows at each state.

Finally, we evaluate the absolute spectral gap $\delta$. The gaps are $0.10158$, $0.27803$, and $0.69251$ for QEMC, IQEMC, and $t$-IQEMC, respectively. Thus, our methods exhibit faster asymptotic convergence per complete Monte Carlo update for this instance. These calculations provide a numerical consistency check of the implementation: the path-based updates preserve the target stationary distribution while violating detailed balance.

\textit{Comparisons with other methods} --  To compare the convergence of IQEMC and $t$-IQEMC across different quantum circuits and under matched quantum resources, we benchmark them against QEMC and QEMC+. QEMC+ feeds each measured output back into the circuit for $n$ successive steps, followed by random selection of one measured state for acceptance or rejection, facilitating exploration across mixing bottlenecks.

We benchmark the four methods by examining the scaling of the spectral gap with system size for the fully connected random Ising model. In Fig.~\ref{fig:End_Matter_result}, both IQEMC and $t$-IQEMC achieve the largest spectral gaps. IQEMC exhibits the most favorable scaling with system size, while $t$-IQEMC preserves this scaling advantage with a larger absolute spectral gap. QEMC+ shows no significant difference from QEMC and remains below IQEMC and $t$-IQEMC in both scaling and absolute spectral gap.

\end{document}


\title{Supplemental Material: Irreversibility Enhances Quantum-Enhanced Markov-Chain Monte Carlo}

\author{Zidong Cui}
\email[Corresponding authors: ]{202411210429@std.uestc.edu.cn}
\altaffiliation{These authors contributed equally}
\affiliation{Institute of Fundamental and Frontier Sciences, University of Electronic Science and Technology of China, Chengdu 611731, China}
\affiliation{School of Physics, University of Electronic Science and Technology of China, Chengdu 611731, China}

\author{Kefan Cao}
\altaffiliation{These authors contributed equally}
\affiliation{Institute of Fundamental and Frontier Sciences, University of Electronic Science and Technology of China, Chengdu 611731, China}
\affiliation{Yingcai Honor College, University of Electronic Science and Technology of China, Chengdu 611731, China}

\author{Jiale Chen}
\affiliation{Institute of Fundamental and Frontier Sciences, University of Electronic Science and Technology of China, Chengdu 611731, China}

\author{Lei Wang}
\affiliation{Beijing National Laboratory for Condensed Matter Physics and Institute of Physics, Chinese Academy of Sciences, Beijing 100190, China}

\author{Ying Tang}
\email[Corresponding authors: ]{jamestang23@gmail.com}
\affiliation{Institute of Fundamental and Frontier Sciences, University of Electronic Science and Technology of China, Chengdu 611731, China}
\affiliation{School of Physics, University of Electronic Science and Technology of China, Chengdu 611731, China}
\affiliation{Key Laboratory of Quantum Physics and Photonic Quantum Information, Ministry of Education, University of Electronic Science and Technology of China, Chengdu 611731, China}
\affiliation{Non-classical Information Science Basic Discipline Research Center of Sichuan Province, University of Electronic Science and Technology of China, Chengdu 611731, China}


\maketitle
\tableofcontents

\clearpage

\section{State encoding}
In this section, we define the index of each state and its corresponding quantum-state encoding, which are used to order the candidate states for probability flow allocation in irreversible quantum-enhanced Monte Carlo (IQEMC). Supplemental Table~\ref{tab:state} summarizes this convention and assigns each state a unique integer label. 
\begin{table}[h]
\centering
\caption{Encoding convention for spin states, computational-basis states, and integer labels.}
\label{tab:state}
\begin{tabular}{ccc}
\hline
\multicolumn{3}{c}{Spin state representation} \\
\hline
State& Encoding & Index\\
$(1,\ldots,1,1)$ & $|0\ldots00\rangle$ & $0$ \\
$(1,\ldots,1,-1)$ & $|0\ldots01\rangle$ & $1$ \\
$(1,\ldots,-1,1)$ & $|0\ldots10\rangle$ & $2$ \\
$\ldots$ & $\ldots$ & $\ldots$ \\
$(-1,\ldots,-1,-1)$ & $|1\ldots11\rangle$ & $2^n - 1$ \\
\hline
\end{tabular}
\end{table}

\section{Roles of inverse circuits, circuit selection, and random anchors}
\label{sec:path-construction-conditions}

This section clarifies the roles of inverse circuits, circuit selection, and random anchors in the auxiliary-path construction. The conditions discussed here apply to this particular construction; they are not necessary conditions for all valid Markov-chain Monte Carlo (MCMC) algorithms. In particular, inverse-circuit sampling provides a common quantum
path factor, whereas random circuit selection is not required for stationarity. Random anchors allow a fixed ordered path to admit multiple possible current states. Preservation of the target distribution alone does not establish ergodicity or a sampling advantage.

\subsection{Cancellation of quantum proposal probabilities}

In this subsection, we generalize the path construction to arbitrary circuit-selection and anchor distributions and show that the quantum transition probabilities cancel from the conditional weights. Let $\mathcal C$ be the finite configuration space, with target distribution $\pi(\mathbf{s})\propto e^{- E(\mathbf{s})/T}$. Consider a fixed bank of state-independent unitary circuits $\{U_k\}_{k=1}^{K}$, with forward proposal probabilities $q_k(\mathbf{s},\mathbf{s}') = \left| \langle\mathbf{s}'|U_k|\mathbf{s}\rangle\right|^2$. We generalize the uniform circuit selection used in the main text to a fixed selection rule $r(k\mid\mathbf{s})$, satisfying
\begin{equation}
    r(k\mid\mathbf{s})\geq0,
    \qquad
    \sum_{k=1}^{K}r(k\mid\mathbf{s})=1.
\end{equation}
The selected circuit index remains fixed throughout the generation of each auxiliary path $\mathbb{Z}=(\cdots)$.	The ordered path is $ \mathbb Z=\bigl(\mathbf{x}_{0},\ldots,\mathbf{x}_{n}\bigr)$, with $n+1$ positions. An anchor position $\xi$ is drawn independently of the current state and circuit index, with probability
\begin{equation}
    \Pr(\xi=z)=p_z,
    \qquad
    p_z\geq0,
    \qquad
    \sum_{z=0}^{n}p_z=1.
\end{equation}
After setting $\mathbf{x}_{\xi}=\mathbf{s}$, we generate the path toward increasing indices using $U_k$ and toward decreasing indices using $U_k^\dagger$. Each circuit application is followed by computational-basis measurement, and the measured state provides the input to the next application. For the auxiliary object $A=(k,\mathbb Z)$, define
\begin{equation}
\left \{
    \begin{aligned}
         B_A&=\prod_{z=0}^{n-1}q_k\bigl(\mathbf{x}_{z},\mathbf{x}_{z+1}\bigr),\\
        \eta_A(\mathbf{s})&=\sum_{z=0}^{n} p_z\mathbf{1}_{\{\mathbf{x}_{z}=\mathbf{s}\}},\\
        c_A(\mathbf{s}) &=\sum_{z=0}^{n}\mathbf{1}_{\{\mathbf{x}_{z}=\mathbf{s}\}}.
    \end{aligned}\right .
\end{equation}

Here $\eta_A(\mathbf{s})$ is the total anchor probability associated with occurrences of $\mathbf{s}$, and $c_A(\mathbf{s})$ is its multiplicity along the path. The anchor is marginalized out when forming the conditional update. The inverse circuit satisfies $\left|\langle\mathbf{s}|U_k^\dagger|\mathbf{s}'\rangle\right|^2=\left|\langle\mathbf{s}'|U_k|\mathbf{s}\rangle\right|^2=q_k(\mathbf{s},\mathbf{s}')$. Consequently, for any anchor $\xi$, the probability of generating a specified ordered path, conditional on $\mathbf{x}_{\xi}=\mathbf{s}$ and $k$, is
\begin{align}
    &\prod_{z=0}^{\xi-1}\left|\langle\mathbf{x}_{z}|U_k^\dagger|\mathbf{x}_{z+1}\rangle\right|^2
\nonumber\times\prod_{z=\xi}^{n-1} \left| \langle\mathbf{x}_{z+1}|U_k|\mathbf{x}_{z}\rangle\right|^2 = B_A,
\end{align}
where empty products equal one. The factor $B_A$ is therefore independent of the anchor position. Summing over anchors gives
\begin{equation}
    p(A\mid\mathbf{s})
    =
    r(k\mid\mathbf{s})\eta_A(\mathbf{s})B_A.
    \label{eq:conditions-generation}
\end{equation}
Let $\mathcal C_A$ denote the set of distinct states on the path, and define $\mu(A)=\sum_{\mathbf{s}\in\mathcal C}\pi(\mathbf{s})p(A\mid\mathbf{s})$. For $\mu(A)>0$, the conditional distribution is
\begin{equation}
\left\{
\begin{aligned}
    \rho_A(\mathbf{s})&=\frac{\pi(\mathbf{s})p(A\mid\mathbf{s})}{\mu(A)}=\frac{ w_A(\mathbf{s})}{ W_A},\\
     w_A(\mathbf{s})&=e^{-E(\mathbf{s})/T}r(k\mid\mathbf{s})\eta_A(\mathbf{s}),\\
     W_A &=\sum_{\mathbf{u}\in\mathcal C_A}w_A(\mathbf{u}).
\end{aligned}
    \right .
    \label{eq:conditions-weights}
\end{equation}
The common quantum factor $B_A$ cancels. Thus, the conditional weights require energies, circuit-selection probabilities, and anchor probabilities, but no explicit evaluation of quantum proposal probabilities. For the uniform choices used in the main text, $r(k\mid\mathbf{s})=1/K$ and $p_z=1/(n+1)$, these weights reduce to $ w_A(\mathbf{s})=\frac{c_A(\mathbf{s})e^{- E(\mathbf{s})/T}}{K(n+1)}$. Dropping the common factor $1/[K(n+1)]$ recovers $w_A(\mathbf{s})=c_A(\mathbf{s})e^{- E(\mathbf{s})/T}$.

\subsection{Circuit randomness and state-dependent selection}

In this subsection, we analyze how the circuit-selection rule affects the conditional weights and connectivity of the resulting Markov chain. In particular, we show that state-independent selection is sufficient for preserving the target distribution, whereas state-dependent selection can lead to reducibility when some selection probabilities vanish. Random circuit selection is not necessary for preserving the target distribution. If the same circuit $U_{k_0}$ is used at every state, then
\begin{equation}
    r(k\mid\mathbf{s})
    =
    \mathbf{1}_{\{k=k_0\}},
\end{equation}
and the conditional weights are proportional to $e^{- E(\mathbf{s})/T}\eta_A(\mathbf{s})$. More generally, any state-independent selection rule $r(k\mid\mathbf{s})=r_0(k)$ contributes only a common factor for a fixed selected index $k$. In contrast, state-dependent selection must be included in the conditional weights. For a deterministic rule $k=f(\mathbf{s})$,
\begin{equation}
     w_A(\mathbf{s}) =e^{-E(\mathbf{s})/T}\eta_A(\mathbf{s})\mathbf{1}_{\{f(\mathbf{s})=k\}}.
    \label{eq:conditions-deterministic}
\end{equation}
A state with $f(\mathbf{s})\neq k$ has zero conditional weight, even if it appears on the generated path. A flow allocation that preserves these weights cannot transfer probability from a state with positive weight to one with zero weight. As a result, the class label $f(\mathbf{s})$ stays the same throughout the chain when this update is used alone. If there is more than one nonempty class, the chain is reducible, even though it still preserves $\pi$. A strictly positive selection rule, for example
\begin{equation}
    r(k\mid\mathbf{s})=\frac{\varepsilon}{K}+(1-\varepsilon)\widetilde{r}(k\mid\mathbf{s}),\qquad 0<\varepsilon\leq1,
\end{equation}
where $\widetilde{r}(k\mid\mathbf{s})\geq0$ and $\sum_{k=1}^{K}\widetilde{r}(k\mid\mathbf{s})=1$, guarantees $r(k\mid\mathbf{s})\geq\varepsilon/K>0$ for every circuit and state, thereby removing this particular obstruction.

This alone does not ensure that the quantum proposals connect all states or improve sampling efficiency. Moreover, selecting circuits based on the current state does not justify changing the circuit index along a path. The adjoint identity pairs $U_k$ with $U_k^\dagger$ for the same $k$.

\subsection{Random anchors and nontrivial conditional updates}

In this subsection, we show that the multiplicity factor arises naturally from marginalizing the anchor and examine how different anchor-selection rules affect the resulting conditional update. Uniform anchor sampling gives
\begin{equation}
    \eta_A(\mathbf{s})
    =
    \frac{c_A(\mathbf{s})}{n+1}.
\end{equation}
Hence, up to a common normalization factor, the conditional weights are $ w_A(\mathbf{s})\propto  c_A(\mathbf{s})e^{- E(\mathbf{s})/T} r(k\mid\mathbf{s})$. The multiplicity factor is therefore a consequence of marginalizing the uniformly sampled anchor, rather than an additional heuristic. If the anchor is instead fixed at $\xi_0$, then
\begin{equation}
    \eta_A(\mathbf{s})
    =
    \mathbf{1}_{\{\mathbf{x}_{\xi_0}=\mathbf{s}\}}.
\end{equation}
For an auxiliary object that can actually be generated, the conditional distribution is consequently $\rho_A(\mathbf{s})=\mathbf{1}_{\{\mathbf{x}_{\xi_0}=\mathbf{s}\}}$. A conditional update preserving this distribution must return $\mathbf{x}_{\xi_0}$, which is the current state. Thus, a fixed anchor makes this particular conditional update trivial.

This conclusion concerns conditioning on the same ordered path. Generating a different path from a proposed new configuration in the next iteration does not establish balance for the current update. Algorithms using fixed anchors can still be valid, but they require a different construction or a separate balance proof. Uniformity itself is not essential. Nonuniform, fixed anchor probabilities are compatible with Eq.~\eqref{eq:conditions-weights}, provided that the factor $\eta_A(\mathbf{s})$ is retained. Choosing $p_z>0$ for every position allows every occurrence on the path to serve as an anchor, but does not guarantee that the path contains distinct configurations or that the marginal chain is irreducible.

\subsection{Conditional flow allocation and global balance}

In this subsection, we establish that the proposed flow allocation preserves the target distribution by satisfying global balance, provided that the conditional weights are consistent with the actual path-generation rule. For each $A$ with $\mu(A)>0$, define the positive-weight candidate set $  \mathcal C_A^+= \left\{\mathbf{s}\in\mathcal C_A: w_A(\mathbf{s})>0 \right\}$. Construct nonnegative conditional flows satisfying
\begin{equation}
    \sum_{\mathbf{s}'\in\mathcal C_A^+} v_A(\mathbf{s},\mathbf{s}')
    =  w_A(\mathbf{s})
    =\sum_{\mathbf{s}'\in\mathcal C_A^+}v_A(\mathbf{s}',\mathbf{s}).
    \label{eq:conditions-flow}
\end{equation}
The probability-flow allocation defined in the main text provides such a construction. The candidate ordering and flow allocation must define a single kernel for each fixed auxiliary object $A$; they must not be recomputed using an uncorrected dependence on the current state or the sampled anchor. For $\mathbf{s}\in\mathcal C_A^+$, let
\begin{equation}
    P_A(\mathbf{s},\mathbf{s}') =\frac{v_A(\mathbf{s},\mathbf{s}')}{ w_A(\mathbf{s})}.
\end{equation}
Zero-weight candidates are excluded, avoiding division by zero, and transitions to configurations outside $\mathcal C_A^+$ are set to zero. Eq.~\eqref{eq:conditions-flow} implies
\begin{align}
    \sum_{\mathbf{s}\in\mathcal C_A^+}\rho_A(\mathbf{s})P_A(\mathbf{s},\mathbf{s}')=\frac{1}{ W_A}\sum_{\mathbf{s}\in\mathcal C_A^+} v_A(\mathbf{s},\mathbf{s}')=\rho_A(\mathbf{s}').
\end{align}
The marginal transition kernel is $P(\mathbf{s}'\mid\mathbf{s}) =\sum_{A:\,p(A\mid\mathbf{s})>0}p(A\mid\mathbf{s})P_A(\mathbf{s},\mathbf{s}')$. It therefore satisfies
\begin{align}
    \sum_{\mathbf{s}\in\mathcal C}\pi(\mathbf{s})P(\mathbf{s}'\mid\mathbf{s})&=\sum_{A:\,\mu(A)>0}\mu(A) \sum_{\mathbf{s}\in\mathcal C_A^+}\rho_A(\mathbf{s})P_A(\mathbf{s},\mathbf{s}')
    \nonumber\\
    &=\sum_{A:\,\mu(A)>0}\mu(A)\rho_A(\mathbf{s}')
    \nonumber\\
    &=\pi(\mathbf{s}')\sum_A p(A\mid\mathbf{s}')=\pi(\mathbf{s}').
\end{align}
This establishes global balance. The flow allocation preserves the conditional distribution derived from the actual path-generation rule; it does not correct incorrect conditional weights. Moreover, asymmetric conditional flows do not automatically imply irreversibility of the marginal chain, because averaging over auxiliary objects can cancel their contributions. Detailed-balance violation must therefore be checked for the resulting kernel $P$. Convergence additionally requires irreducibility and aperiodicity, and any sampling advantage must be established separately.

\section{Comparison with Classical MCMC}
\begin{figure}[htbp]
    \centering
    \includegraphics[width=0.5\linewidth]{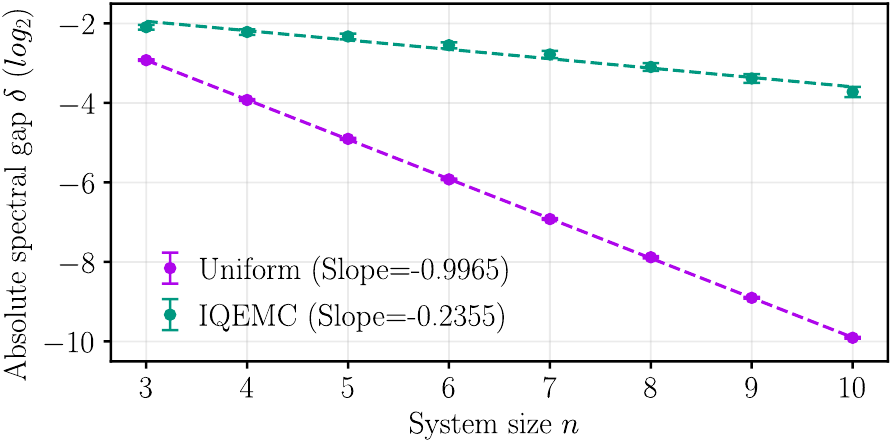}
    \caption{Average absolute spectral gap of IQEMC and classical MCMC with uniform proposals in $500$ fully connected random Ising models ($T=0.1$). The spectral-gap advantage of IQEMC over classical MCMC with uniform proposals increases with system size, indicating a polynomial scaling advantage.}
    \label{fig:sup_result}
\end{figure}

In this section, we compare IQEMC with classical MCMC algorithms using uniform proposals. We benchmark the spectral gap $\delta$ to demonstrate the efficiency of IQEMC compared with classical MCMC with uniform proposals in the fully connected random Ising model:
\begin{align}
    H_{\text{prob}}=-\sum_{l<m} J_{lm} Z_l Z_m - \sum_l h_l Z_l,
\end{align}
where $J_{lm}$ and $h_l$ are drawn from standard normal distributions. For system sizes $3\leq n\leq 10$, IQEMC achieves a larger $\delta$ than classical MCMC with uniform proposals, demonstrating its advantage over classical MCMC (Supplemental Figure~\ref{fig:sup_result}). Results are averaged over $500$ instances.

\begin{figure}[htbp]
\centering
{\includegraphics[width=1\textwidth]{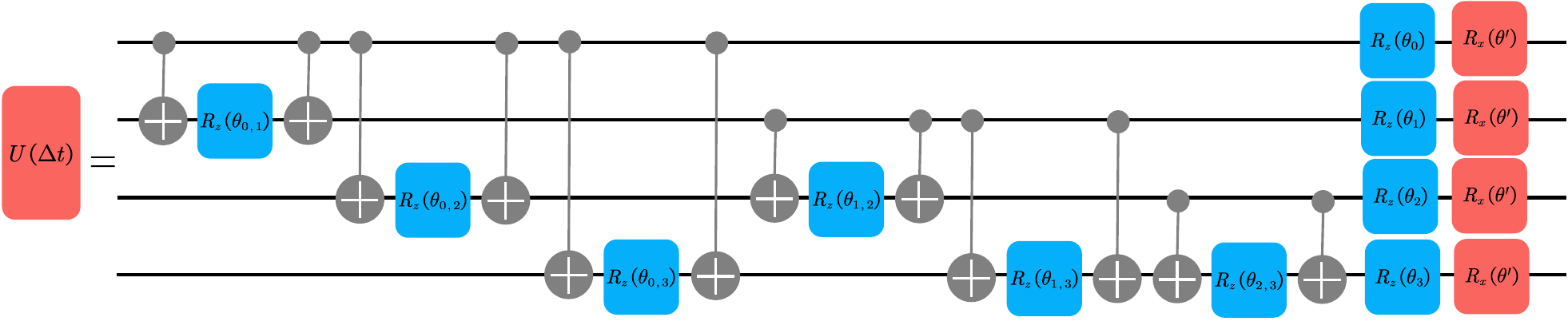}}
\caption{First-order Trotter circuit approximating one short-time evolution step $\exp[-iH_k(t_\nu)\Delta t]$. The mixing parameter is constant, $\gamma_k(t_\nu)=\gamma_k$, for IQEMC and follows $\gamma_k(t_\nu)=\gamma_{0,k}+\kappa_k t_\nu$ for $t$-IQEMC. The $R_z$ gates encode local fields $h_l$, CNOT-$R_z$-CNOT blocks encode pairwise couplings $J_{lm}$, and $R_x$ gates implement the mixing Hamiltonian $H_{\mathrm{mix}}$.
}
\label{fig3}
\end{figure}
\section{Quantum Circuit}

In this section, we describe how the circuits used in IQEMC and $t$-IQEMC can be decomposed into elementary quantum gates. Each circuit is specified by its index $k$, and its parameters are independent of the input configuration.

\subsection{First-Order Trotter Circuits}

Following the End Matter, the Hamiltonian is
\begin{equation}
H_k(\tau)=[1-\gamma_k(\tau)]H_{\mathrm{prob}}+\gamma_k(\tau)H_{\mathrm{mix}},
\end{equation}
where
\begin{equation}
H_{\mathrm{prob}}=-\sum_{l<m}J_{lm}Z_lZ_m-\sum_l h_lZ_l,\qquad H_{\mathrm{mix}}=\sum_l X_l.
\end{equation}
For IQEMC, $\gamma_k(\tau)=\gamma_k$ is constant. For $t$-IQEMC, $\gamma_k(\tau)=\gamma_{0,k}+\kappa_k\tau$, where $\kappa_k$ remains fixed during each circuit. The circuit parameters and evolution intervals are specified in the End Matter. We set $\hbar=1$.

For a selected circuit $k$, let $\Delta t$ be the duration of each retained interval, $t_\nu$ its midpoint, and $N_{\mathrm{seg}}$ the number of intervals. The midpoint approximation and first-order Trotter decomposition give
\begin{equation}
e^{-iH_k(t_\nu)\Delta t}\approx V_{k,\nu}^{(1)}=e^{-i\gamma_k(t_\nu)H_{\mathrm{mix}}\Delta t}e^{-i[1-\gamma_k(t_\nu)]H_{\mathrm{prob}}\Delta t}.
\end{equation}
Using $R_\mu(\theta)=e^{-i\theta\sigma_\mu/2}$ and $R_{ZZ}^{(lm)}(\theta)=e^{-i\theta Z_lZ_m/2}$, the two blocks are
\begin{align}
e^{-i\gamma_k(t_\nu)H_{\mathrm{mix}}\Delta t}
&=\prod_l R_x^{(l)}\!\left(2\gamma_k(t_\nu)\Delta t\right),\\
e^{-i[1-\gamma_k(t_\nu)]H_{\mathrm{prob}}\Delta t}
&=\prod_{l<m}R_{ZZ}^{(lm)}\!\left(-2[1-\gamma_k(t_\nu)]J_{lm}\Delta t\right)\nonumber\\
&\quad\times\prod_l R_z^{(l)}\!\left(-2[1-\gamma_k(t_\nu)]h_l\Delta t\right).
\end{align}
All terms in $H_{\mathrm{prob}}$ commute, so the second identity is exact. Each interaction gate is implemented using
\begin{equation}
R_{ZZ}^{(lm)}(\theta)=\operatorname{CNOT}_{l\rightarrow m}R_z^{(m)}(\theta)\operatorname{CNOT}_{l\rightarrow m}.
\end{equation}
These elementary gates form the circuit blocks shown in Supplemental Figure~\ref{fig3}, which can all be decomposed into standard gate primitives supported by current quantum hardware~\cite{javadi2024quantum}. The complete first-order circuit is
\begin{equation}
U_k^{(1)}=V_{k,N_{\mathrm{seg}}}^{(1)}\cdots V_{k,1}^{(1)},
\end{equation}
where the rightmost factor acts first.

\subsection{Second-Order Trotter Circuits}

A second-order implementation uses a symmetric decomposition into the same problem and mixing blocks:
\begin{equation}
\begin{aligned}
V_{k,\nu}^{(2)}
={}&e^{-i[1-\gamma_k(t_\nu)]H_{\mathrm{prob}}\Delta t/2}\\
&\times e^{-i\gamma_k(t_\nu)H_{\mathrm{mix}}\Delta t}\\
&\times e^{-i[1-\gamma_k(t_\nu)]H_{\mathrm{prob}}\Delta t/2}.
\end{aligned}
\end{equation}
Each problem half-step uses the same $R_z$ and $R_{ZZ}$ gates as above with their angles halved, while the mixing step retains the full $R_x$ rotation angles. The complete second-order circuit is
\begin{equation}
U_k^{(2)}=V_{k,N_{\mathrm{seg}}}^{(2)}\cdots V_{k,1}^{(2)}.
\end{equation}
Thus, the second-order circuit requires no additional types of elementary gates.

\subsection{Inverse Circuits and Candidate Paths}

At either Trotter order, the inverse circuit is obtained by reversing the gate sequence and taking the adjoint of each gate. Rotation angles change sign, whereas CNOT gates remain unchanged. Denoting the implemented forward circuit by $U_k$, this construction ensures
\begin{equation}
|\langle\mathbf{s}'|U_k|\mathbf{s}\rangle|^2=|\langle\mathbf{s}|U_k^\dagger|\mathbf{s}'\rangle|^2
\end{equation}
for ideal gate operations. This identity supplies the forward--inverse reciprocity required by the global-balance proof, even when the forward evolution is Trotterized.

Each candidate-path link applies one complete forward or inverse circuit, followed by measurement and re-preparation of the measured configuration for the next link. The Trotter segments within each circuit are executed coherently, without intermediate measurements.

\bibliography{bib}